\documentclass[10pt, conference]{IEEEtran}
\IEEEoverridecommandlockouts

\ifCLASSINFOpdf
\else
\fi

\usepackage{graphicx, epstopdf}
\usepackage{multirow}
\usepackage{threeparttable}

\usepackage{afterpage}

\usepackage{comment}

\usepackage{cite}
 
\usepackage{amsmath}

\usepackage{xspace} % Defines a 'smart space' (not present before punctuation)

\newcommand{\fig}{Fig.\xspace}

\newcommand{\ubar}[1]{\mkern3mu\underline{\mkern-3mu #1\mkern-3mu}\mkern3mu}
\newcommand{\obar}[1]{\mkern3mu\overline{\mkern-3mu #1\mkern-3mu}\mkern3mu}

\usepackage[percent]{overpic}

\usepackage{xcolor}
\usepackage{pict2e}

\makeatletter
\newcommand{\removelatexerror}{\let\@latex@error\@gobble}
\makeatother

\usepackage[linesnumbered,lined,boxed,commentsnumbered, noend, ruled]{algorithm2e}

\usepackage{setspace}

\begin{document}

\title{Stochastic Battery Model for Aggregation of Thermostatically Controlled Loads}

\author{\IEEEauthorblockN{Sohail Khan, Mohsin Shahzad, Usman Habib}
\IEEEauthorblockA{AIT Austrian Institute of Technology, Vienna, Austria\\
{sohail.khan@ait.ac.at, \{givenname.surname\}@ait.ac.at}}
%\{givenname.surname\}@ait.ac.at
%mohsin.shahzad@ait.ac.at, usman.habib@ait.ac.at
\and
\IEEEauthorblockN{Wolfgang Gawlik}
\IEEEauthorblockA{TU Vienna -- Austria\\
{gawlik@ea.tuwien.ac.at}}
\and
\IEEEauthorblockN{Peter Palensky}
\IEEEauthorblockA{TU Delft -- Netherlands\\
{p.palensky@tudelft.nl}}
}

\maketitle

\begin{abstract}
%The potential of demand side as a frequency reserve proposes interesting opportunity in handling imbalances due to intermittent renewable energy sources.
%This paper proposes a stochastic battery model for the non-disruptive control of the aggregation of Thermostatically Controlled Loads (TCLs).
%In this method, each TCL sends its status, availability and relative temperature distance till the switching boundary to the central control.
%This information is used to predict the stochastic values of admissible ramp-rate and the state of charge limits of the equivalent battery model.
%The approach builds on and improves the existing research work by providing a straight-forward mechanism of handling the uncertainty. 
%Effectiveness of the proposed approach is demonstrated by a test case having a large number of residential TCLs tracking a real scaled down frequency regulation signal.
The potential of demand side as a frequency reserve proposes interesting opportunity in handling imbalances due to intermittent renewable energy sources.
This paper proposes a novel approach for computing the parameters of a stochastic battery model representing the aggregation of Thermostatically Controlled Loads (TCLs).
A hysteresis based non-disruptive control is used using priority stack algorithm to track the reference regulation signal.
The parameters of admissible ramp-rate and the charge limits of the battery are dynamically calculated using the information from TCLs that is the status (on/off), availability and relative temperature distance till the switching boundary.
The approach builds on and improves on the existing research work by providing a straight-forward mechanism for calculation of stochastic parameters of equivalent battery model.
%of handling the uncertainty in the availability of this resource. 
The effectiveness of proposed approach is demonstrated by a test case having a large number of residential TCLs tracking a scaled down real frequency regulation signal.
\end{abstract}


% \fi
\IEEEpeerreviewmaketitle

\begin{spacing}{1.2}
\section*{Nomenclature}
\addcontentsline{toc}{section}{Nomenclature}
%\subsection{Variables}
\begin{IEEEdescription}[\IEEEusemathlabelsep\IEEEsetlabelwidth{$V_1,V_2$}]
\item[$i$] Index of the TCL, from 1 to $N$.
\item[$k$] Time index, from 1 to $T$.
\item[$\theta^i$] Temperature of $i^{\textrm{th}}$ TCL.
\item[$\Delta^i$] Temperature dead-band width of $i^{\textrm{th}}$ TCL.
\item[$\psi$] Aggregate power deviation from the base value.
\item[$\delta$] Operational status of TCL $\in(0,\;1)$.
\item[$\rho^i$] Time duration after status change of $i^{\textrm{th}}$ TCL.
\item[$\obar{\rho^i}$] Short cycling time constraint of $i^{\textrm{th}}$ TCL.
\item[$\lambda^i$] TCL availability status $\in(0,\;1)$.
\item[$\pi^i$] Normalized temperature distance to the switching boundary of $i^{\textrm{th}}$ TCL.
\item[$P^i,\; P^i_0$] Rated and nominal power of $i^{\textrm{th}}$ TCL.
\end{IEEEdescription}
\end{spacing}

\section{Introduction}
The balance between supply and demand is the principal control objective in the power system.
In order to achieve this balance the planning and control is performed at various time scales ranging from day-ahead to seconds.
The rise in distributed generation from renewable energy sources like wind and solar have increased the uncertainty at generation-side and thus, the imbalance probability~\cite{4808228}.
Generally, the balance is achieved by activating the reserve power from the online generators.
This approach have proved to be effective in power system.
However, increase in the level and volatility of the imbalance have made it an expensive choice.
Thus the alternate sources of flexibility are highly sought after. 
Among them, a promising alternative is the active control of Thermostatically Controlled Loads (TCLs) e.g., air conditioning and heating units~\cite{1318707}.
In these loads, there exists a flexibility around the user set-point called the dead-band. 
A non-disruptive operation can be achieved by controlling the operational state of the TCL within the dead-band in order to track a reference active power signal~\cite{6669582}.
Such that, the load can be increased or decreased without causing discomfort to the customers.

\subsection{TCL as frequency reserve}
The imbalance in supply and demand of active power results in frequency deviation from the nominal value.
Frequency regulation is a real time objective and is critical for a secure operation of the system.
The time scale requirements ranges from seconds to minutes and thus limits the maximum admissible delay in the response of reserve resources.
The aggregation of residential TCLs can be a cost effective and a secure alternative to the fast generation units or storage solutions~\cite{6832599}.
The ability of the TCL aggregation in responding to the real-time frequency regulation signal has been reported in~\cite{koch2011modeling, 6858734}.
The impact of using this resource has been focused in several regional studies. 
Such as, in~\cite{7302276}, the capability of using TCLs as a storage capacity in Switzerland is discussed.
The potential is interesting but the infrastructure required at the TCL level to enable this technology is a limiting factor.
This paper aims to present a cost effective approach for assessing the availability and the dynamic capability of this resource as stochastic parameters of an equivalent battery model.

\subsection{Literature survey}
The aggregate behavior of large population of TCLs stems out from the modeling approaches used.
In literature, two approaches are generally used for modeling TCLs.
The individual load model of the TCL as a combination of a continuous temperature state and a discrete ``switching state'' were first presented 
%by Ihara and Schweppe
in~\cite{4111111}.
Authors in \cite{molina2003implementation} have verified this model for the real population of TCLs. 
%by Molina et al.
%in~\cite{molina2003implementation}.
The three state model capturing the temperature of thermal mass is  discussed 
%by Zhang et al.
in~\cite{6545388}.
More advance models are discussed in literature that aims to model the dynamics of TCL accurately~\cite{5762648}.
The simulation of individual models can be challenging for large number of TCLS.
However, this approach is suited for the simple control strategies~\cite{koch2009active}.
The second approach is the direct modeling of the aggregation of TCLs.
Recently, state space models have been explored with much interest as it facilitates the control design.
Among such methods, partial differential equation based approach is used 
%by Bashash and Fathy 
for designing a sliding model control approach~\cite{6239581}.
The representation of the TCL aggregation by a generalized battery and the calculation of its parameters are interesting research areas and depend on the modeling and control strategies discussed above.

\begin{figure} [htb]
\centering
% left, bottom, right and top 
\includegraphics[trim=1.2cm 0cm 0cm 0.5cm, width=0.45\textwidth]{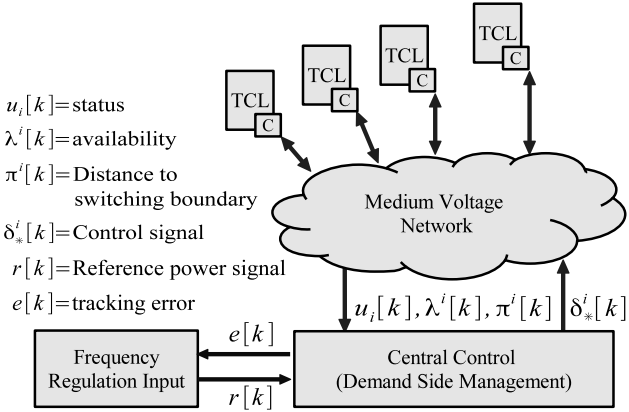}
\caption{Overview of the control mechanism}
\label{figure:system overview}
\end{figure}

The limits on the power and energy capacity that can be tracked by TCL aggregation are discussed in~\cite{6669582}.
The values are calculated as function of the outdoor temperature. 
The other factors affecting the availability like customer preferences are not considered. 
%Hao et al.~
Authors in~\cite{6736573} have recently presented a generalized battery model.
The power limits on this model are derived using continuous power model. 
Similarly, 
%Sanandaji et al.~
authors in~\cite{7345604} have modeled the stochastic parameters of the battery model.
In these papers, the power consumption at each TCL is required to be measured and sent to the central control.
It can be an additional expense along with the consideration of measurement uncertainty in the continuous variable. 
The dynamic battery parameters are obtained from the historical information of the switching status of TCLs.
Alternatively, a novel mechanism of computing the stochastic battery parameters is proposed here.
These parameters are the maximum ramp-up/down values and the charging/discharging potential of the battery.
Analytic expressions are provided to calculate these parameters using status $(u^i[k])$, availability $(\lambda^i[k])$ and relative temperature distance to switching boundary $(\pi^i[k])$. 
Fig.~\ref{figure:system overview} shows the working principle of the TCLs aggregation control. The regulation signal $r[k]$ is the input to the central control which communicates with the TCLs and control their operational states. The error in the tracking signal is sent back for the monitoring purpose. 

%An additional variable of the availability as a binary variable is introduced. 
%The rated power of each TCL is assumed to be known as part of contract with the central control. 
%This approach facilitates the computation of stochastic battery limits at the central level.
%The main contributions are summarized as,
%\begin{itemize}
%\item A mechanism where each TCL transmits the status $u^i[k]$, relative temperature distance to switching boundary~$\pi^i[k]$  and the availability $\lambda^i[k]$ to the central control. This data is used to compute the stochastic battery parameters.
%\item A formulation that computes the stochastic limits of the State of Charge (SoC) and the potential of the aggregation to track the regulation signal. 
%%\item The control performance is analyzed in the presence of communication delay and uncertainty in the short-cycling constraints.
%\end{itemize}

Rest of the paper is organized as follows. 
The model of TCL and stochastic battery is discussed in Sec.~\ref{sec:problem statement}. 
The implementation strategy of the stochastic limits based control is presented in Sec.~\ref{sec:control scheme}. 
In Sec.~\ref{sec:results}, the test case and results are discussed and followed by the conclusion in Sec.~\ref{sec:conclusion}.

\section{Stochastic Battery Model}\label{sec:problem statement}

\subsection{Thermostatically Controlled Loads}\label{sec:thermostatically controlled loads}
In order to simulate the aggregation of TCL, a simplified first order model has been reported widely in the literature~\cite{4111111,6669582,49112}.
It is given by a difference equation,
\begin{equation}
\theta^i[k+1]=g^i\theta^i[k]+(1-g^i)(\theta^i_a[k]-\delta^i[k]\theta^i_g)+\epsilon^i[k]
\enspace , 
\label{eq:TCL model}
\end{equation}
where, $g^i = e^{-h/(R^iC^i)}$ ($h$ is the sampling time and $R^iC^i$ is the time constant of the TCL), $\theta^i_g = R^iP^i\eta^i$ and $\theta^i_a[k]$ is the ambient temperature measurement at $i^{th}$ TCL. The first term represents the decaying influence of the temperature in the previous time step. While, the second term is temperature gain/loss as the TCL is switch On/Off. 
$P^i$ is the rated power of TCL which is positive in case of the air-conditioning load and negative for the heater.
The parameter $\epsilon^i$ represents the noise associated with the temperature measurement of $i^{\textrm{th}}$ TCL. 
The detail of other parameters can be found in Tab.~\ref{table:PV data}.
The state transition of the cooling TCL is given as~\cite{6669582},
\begin{equation}
\delta^i[k+1]=
\left\{\begin{matrix}
1 &\theta^i[k+1] > \theta^i_\textrm{ref}+\Delta^i \\ 
0 &\theta^i[k+1] < \theta^i_\textrm{ref}-\Delta^i \\ 
\delta^i[k] &otherwise
\end{matrix}\right.
\enspace . 
\end{equation}
Here, $\Delta^i$ defines the dead-band around the reference set-point~$\theta^i_{\textrm{ref}}$. When TCL is operating outside the dead-band it is considered non-controllable. 

In order to perform the non-disruptive load control, TCLs are actively controlled while operating within the dead-band~\cite{6669582}.
In the steady state it is assumed that $\theta^i=\theta^i_{\textrm{ref}}$. 
The corresponding power consumption is obtained by solving the continuous power model from~\cite{6832599}, leading to,
\begin{equation}
P_i^o[k] = 
\frac{\theta^i_a[k]-\theta^i_r}{\eta^iR^i}
\enspace .
\end{equation}
It is reported in~\cite{6669582}, that $P_i^o[k]$ can be considered as average power consumed by the TCL operating in the steady state.
The baseline power consumption of the TCL aggregation is given as,
\begin{equation}
P_{\textrm{base}}[k] = \sum_i P_o^i
\enspace .
\end{equation}
The instantaneous active power consumption at $k$ is the sum of the rated power of all active TCLs. It is given as,
\begin{equation}
P_{\textrm{agg}}[k] = \sum_i \delta^i[k]P^i
\enspace .
\end{equation}
The difference between the aggregate and baseline power consumption is given as,
\begin{equation}
\psi[k] =  
P_{\textrm{agg}}[k] - P_{\textrm{base}}[k]
\enspace .
\end{equation}
In this study the air-conditioning TCLs are considered. Thus, $\psi[k]>0$ when the average temperature of the TCL is below the set-point.
It indicates the natural phenomenon of temperature decrease when TCL is active.
In order to use TCLs as a frequency reserve, the reference power imbalance signal, $r[k]$, must be followed in real-time.
This can be achieved by adjusting the number of active TCLs in the aggregation in real-time.
If the TCLs are collectively modeled by a battery, then the reference signal can charge or discharge the battery when $r[k] > \psi[k]$ and $r[t] < \psi[k]$ respectively.
The charging process shall turn On while turning Off TCLs is refered as discharging process.
%Maximum charging the battery shall turn Off all the TCLs while the discharge process will force active state of each TCL.

\subsection{Stochastic Battery Model Parameters} \label{sec: stochastic battery model}
The capacity and maximum charge/discharge rates of a stochastic battery model are given as~\cite{6832599},
\begin{equation}
\begin{split}
C
&=
\sum_i
\left(
1+
\left | 1-\frac{a_i}{\alpha} \right |
\right)
\frac{\Delta_i}{b_i}\\
R_+
&=
\sum_i 
\left(
P_i-P_i^o
\right)\\
R_-
&=
\sum_iP_i^o
\end{split}
\enspace ,
\label{eq:deterministic battery model}
\end{equation}
here,
$a_i=1/(R_iC_i)$, 
$b_i=\eta_i/C_i$ and $d = 
1/N
\sum_i
1/(R_iC_i)$.
This formulation enables the calculation of the maximum capacity [kWh] available in the system as a stochastic variable. 
However, the availability of the TCL is a limiting factor in this regard.

\begin{figure} [htb]
\centering
% left, bottom, right and top 
\includegraphics[trim=0cm 0.2cm 0cm 0.5cm, width=0.3\textwidth]{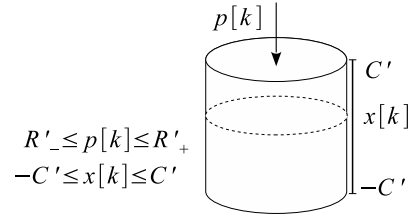}
\caption{Stochastic Battery Model}
\label{figure:battery model}
\end{figure}

The consideration of the availability of TCLs results in a dynamic stochastic model of the battery parameters.
The are formulated as,
\begin{equation}
\begin{split}
C'
&=
\sum_i
\lambda^i[k]
\left(
1+
\left | 1-\frac{a_i}{\alpha} \right |
\right)
\frac{\Delta_i}{b_i}\\
R'_+
&=
R_+
-
\sum_i 
(1-\lambda^i[k])
P_i\\
R'_-
&=
R_-
+
\sum_i 
(1-\lambda^i[k])
P_i
\end{split}
\enspace ,
\label{eq:stochastic battery model}
\end{equation}
where, the $\lambda^i[k]$ is the availability of $i^{\textrm{\textrm{th}}}$ TCL given as,
\begin{equation}
\lambda^i[k]=
\left\{\begin{matrix}
1 & \rho^i[k]>\obar{\rho^i} \;\&\; \ubar{\theta^i}\leq\theta^i[k]\leq\obar{\theta^i}\\ 
0 & \textrm{otherwise}
\end{matrix}\right.
\enspace .
\end{equation}
The stochastic ramp-limit constraint on the reference signal is considered by the following equation,
\begin{equation}
R'_- \leq r[k] \leq R'_+
\enspace . 
\label{eq:stochastic limit constraint}
\end{equation}
While the state of the charge constraint is implemented as,
\begin{equation}
-C' \leq \left( x[k] = \sum_{i}\left[ \frac{\theta^i_{\textrm{ref}}-\theta^i[k]}{b_i}\right]\right) \leq C'
\enspace . 
\label{eq:stochastic charging constraint}
\end{equation}
\textit{Note:} The temperature difference $(\theta^i_{ref}-\theta^i[k])$ in Eq.~\ref{eq:stochastic charging constraint} represents the temperature decrease when the AC is On. Equivalently, it represents the charging of the battery. 
Stochastic battery model is shown in Fig.~\ref{figure:battery model}. The dynamics of the variables are bounded by the stochastic limits and are given as,
\begin{equation} \label{eq1}
\begin{gathered}
%u[k]=\int_o^t
%\psi[k]
%\quad \forall t>0 \;\; \\
\dot{x}[k] = -d x[k] - p[k], \; x(0)=0, \; 
\left | x[k] \right | \leq C'
\quad
%\forall t>0 
\end{gathered}
\enspace ,
\end{equation}
where, the dissipation rate ($d$) is given as, 
\begin{equation}
d = 
\frac{1}{N}
\sum_{k=1}^N
\frac{1}{R_iC_i}
\enspace .
\end{equation}

\section{Control Scheme}\label{sec:control scheme}
\subsection{Hysteresis Based Control}
A hysteresis based control approach is used in this work. 
In this approach, each TCL communicates its operational status, availability and temperature distance information to the central control.
The relative temperature distance $\pi^i[k]$ of a TCL from its switching boundary is used as a priority variable. 
The central control formulate the merit order to turn On/Off the units based on the least distance to the switching boundary.
This approach decreases the cycling of TCL units.
The frequency regulation signal from an energy market is considered as a reference input. 
This real time signal is used to change the state of $i^{\textrm{th}}$ TCL corresponding to the regulation requirements.
The condition when $\psi[k]$ is equal to $r[k]$ indicates the successful tracking.
When $r[k]>\psi[k]$, 
the algorithm at central control turns sufficient number of available units ON.
This step is termed as charging of the battery.
In case of AC loads the impact will be decrease in the average TCL temperature from the set-point value.
Similarly, in converse case sufficient number of On units are turned Off to follow the regulation signal.
The operational mechanism at the central control is shown in Alg.~\ref{algo: central control}.

%The level of TCL specification information requirement at the central control can be significantly reduced if the stochastic energy state prediction is not required.
%In this case, the only values required is the ratings of the TCLs.
%It is assumed that the central control has this information. 
%This value can be the part of the contract of the TCL with the central control.
The algorithm facilitates the addition/removal of the TCLs in the network dynamically.
This can be easily captured during the update of the stochastic battery limits in Eq.~\ref{eq:stochastic battery model}.
The state transitions $\delta^i_*[k]$ triggered by the central control are communicated to the TCLs using the communication network.
At individual TCL level, a local control strategy is applied to respond to this input.
The control algorithm at the TCL is outlined in Alg.~\ref{algo: TCL control}.
The evolution of the TCL internal temperature is modeled by Eq.~\ref{eq:TCL model}.
The TCL is available if the temperature lies in the dead-band region and the short cycling constraint is fulfilled.
The short cycling constraint implies the minimum duration of time that a TCL must remain in a state after a state transition.
The cycling is represented by $\rho^i[k]$.
The violation of this constraint or the operation of the TCL outside the dead-band implies the non-availability.
The temperature distance of the TCL from the switching boundary is normalized with respect to the dead-band width before transmitted to the central control. 
%This is the primary input for the operation of priority based central control.
%The information about the availability, operational status and relative temperature distance is communicated by each TCL to the central control at the central time step.
%The detailed mechanism is outlined in Alg.~\ref{algo: TCL control}.

%(\emph{(xxxx)})

%\begin{figure}[htb]
%%\caption{This algorithm models the forecast error in a stochastic variable considering the spatio-temporal information.}
%\removelatexerror
%\begin{algorithm}[H]
%\label{algo: TCL control}
%\caption{Algorithm at $i^{\textrm{th}}$ TCL}
%Get control signal information from control center $\delta^i_*[k]$\;
%\If {$\rho^i[k+1]>\obar{\rho^i}$ \& $\lambda^i[k]\;=\;1$}
%	{$\delta^i[k] = \delta^i_*[k]$; (Acceptance of central control signal)}
%Calculate $\theta^i[k+1]$ using \equ~\ref{eq:TCL model}\;
%\eIf {$\ubar{\theta^i} \leq \theta^i[k+1] \leq \obar{\theta^i}$}
%{$\delta^i[k+1]=\delta^i[k]$\; 
%$\lambda^i[k+1]=1$\;
%$\rho^i(t+1)=\rho^i[k]+1$
%}
%%\Else
%{
%%$\theta^i[k+1]=1$ \; 
%\lIf {$\theta^i[k+1]>\obar{\theta^i}$} {$\lambda^i[k+1]=1,\; \rho^i[k+1]=0$}
%\lIf {$\theta^i[k+1]<\ubar{\theta^i}$}{$\lambda^i[k+1]=0,\; \rho^i[k+1]=0$}
%}
%\lIf{$\delta^i[k+1]=1$} 
%{$\pi^i[k]\; = \; \left(\theta^i[k+1]-\ubar{\theta^i}\right)/\Delta^i$}
%\lIf{$\delta^i[k+1]=0$} 
%{$\pi^i[k]\; = \; \left(\ubar{\theta^i}-\theta^i[k+1]\right)/\Delta^i$}
%\If {$\rho^i[k+1]\leq\obar{\rho^i}$}
%{
%$\lambda^i[k+1]\;=\ 0$; (Enforce short cycle constraint)
%}
%Transmit $(u_i[k+1],\;\lambda^i[k+1],\;\pi^i[k+1])$ 
%\end{algorithm}
%\end{figure}

\begin{figure}
%\caption{This algorithm models the forecast error in a stochastic variable considering the spatio-temporal information.}
\removelatexerror
\begin{algorithm}[H]
\label{algo: central control}
\caption{Algorithm at Main Control}
\SetKwInOut{Input}{Input}\SetKwInOut{Output}{Output}
\Input{TCL $i$ data $(u_i(t),\;\lambda^i(t),\;\pi^i(t))$}
\Output{Forced state of the TCL $\delta^i_\ast$}
\BlankLine
\emph{Calculate battery parameters $(C,\;n_+,\;n_-)$}\; 
\For( \emph{(Time iteration loop)}){$t := 1 \cdots T$}
   {
      Sample input frequency regulation signal $r(t)$\;
      \For( \emph{(TCL iteration loop)}){$i := 1 \cdots N$}
   {
      Sort priority list of available On/Off TCLs\;
   }
   }
\emph{Update stochastic battery limits $(C',\;n'_+,\;n'_-)$}\;
     \If{$(R'_+ \leq r(t) \leq R'_- )$}
     	{
     	$\xi = r(t)-\psi(t)$ \;
     	\If(\emph{(Priority list based control)}){$r(t) < \psi(t)$}
     		{Turn Off available TCLs till $\delta P < \xi$\;}
     	\Else
     		{Turn On available TCLs till $\delta P < -\xi$\;}
     	}
     \Else
     	{\emph{Regulation not possible}\;}
\end{algorithm}
\vspace{-10pt}
\end{figure}
\begin{figure}
\removelatexerror
\begin{algorithm}[H]
\label{algo: TCL control}
\caption{Algorithm at TCL}
\KwIn{Control signal $\delta^i_*(t)$}
\KwOut{$(u_i(t+1),\;\lambda^i(t+1),\;\pi^i(t+1))$ }
$(\theta^i(t+1), \delta^i(t+1), \lambda^i(t+1))= 0$\;
\If {$\rho^i(t)>\obar{\rho^i}$}
	{$\lambda^i(t+1)\;=\;1$\;
	\If {$\delta^i_*(t)$ is received}
	{$\delta^i(t) = \delta^i_*(t)$\;
	 $\rho^i(t)=0$\;}
	 }
$\rho^i(t+1)=\rho^i(t)+1$\;
%$\theta^i(t+1)=g^i\theta^i(t)+(1-g^i)(\theta^i_a(t)-$\\
%\qquad \qquad \quad $\delta^i(t)\theta^i_g)+\epsilon^i(t)$\;
$\theta^i(t+1)=g^i\theta^i(t)+(1-g^i)(\theta^i_a(t)-
\delta^i(t)\theta^i_g)+\epsilon^i(t)$\;
\If {
$\ubar{\theta^i} \leq \theta^i(t+1) \leq \obar{\theta^i}$}
{$\delta^i(t+1)=\delta^i(t)$\; 
}
\Else
{
$\lambda^i(t+1)=0$\;
\If {$\theta^i(t+1)<\ubar{\theta^i}$}
{
$\delta^i(t+1)=0$\; 
\If {$\delta^i(t)=1$}
{
$\rho^i(t+1)=0$
}
}
\If {$\theta^i(t+1)>\ubar{\theta^i}$}
{
$\delta^i(t+1)=1$\; 
\If {$\delta^i(t)=0$}
{
$\rho^i(t+1)=0$
}
}
}
\lIf{$\delta^i(t+1)=1$} 
{$\pi^i(t)\; = \; \left(\theta^i(t+1)-\ubar{\theta^i}\right)/\Delta^i$}
\lIf{$\delta^i(t+1)=0$} 
{$\pi^i(t)\; = \; \left(\ubar{\theta^i}-\theta^i(t+1)\right)/\Delta^i$}
\end{algorithm}
\vspace{-10pt}
\end{figure}

%\subsection{Stochastic limits consideration}\label{sec:Stochastic limits consideration during operation}
The real time reference signal is compared to the stochastic limits from Eq.~\ref{eq:stochastic limit constraint}.
These limits can be used as a filter to guarantee the tracking of the reference signal.
If violated, the residual can be allocated to the other resources in the network like spinning reserves and storage.
Furthermore, the methodology provides a stochastic energy state estimation of the TCLs aggregation.

\subsection{Regulatory requirements}
Each system operator has specific regulatory requirements for enable the participation of demand side in the frequency regulation process.
For example, CAISO has defined the non-generator resources to provide power bid on the basis of their 15-minute energy capacity~\cite{6858734}. 
There are strict requirements on the telemetry of the TCL data.
In this case, TCLs are required to update their state of charge and instantaneous power status every 4 seconds.
The minimum resource size restriction also limits the potential of this resource. 
CAISO defines minimum resource size of 0.5 MW for a TCL.
This paper target small scale TCLs and the regulatory requirements are considered during the control design.
It compliments the findings proposed in~\cite{7345604} that the control of large number of residential TCLs can provide a reliable regulation service if the uncertainty in the availability are inherently taken into account during the operation.
% if the parameters of stochastic battery parameters are computed.
%In the proposed mode, the aggregator (central control) sends the stochastic values of SoC and ramp limits to the operator in real time.

\section{Results}\label{sec:results}
\subsection{Experimental setup}
The aggregation is composed of $1000$ TCLs.
The specifications are obtained using the table~\ref{table:PV data}.
The parameters of each TCL are obtained by sampling the normal distribution around the values. 
Here, the limits are taken as the percentage deviation from the mentioned quantity.
These limits can be controlled to alter the heterogeneity in the aggregation. 
The heterogeneity of $30 \%$ is considered.
The TCL model time step is $10.02\;sec$.
\begin{table}[htp]
\begin{center}
\caption{Parameters of a typical residential AC TCL \cite{6832599}}
\label{table:PV data}
\renewcommand{\arraystretch}{1.1}
    \begin{tabular}{ | p{1.3cm} | p{3.8cm}| p{0.9cm} | p{1.1cm}| } 
    \hline
    Parameter & Description  & Value & Unit  \\ \hline
    \hline
	$C^i$ & Thermal capacitance & 2 & kWh/$^{\circ}$C		\\ 
	%\hline
	$R^i$ & Thermal resistance & 2 & $^{\circ}$C/kW\\ 		%\hline
	$\theta^i_{ref}$ & Temperature & 22.5 & $^{\circ}$C		\\ 
	%\hline
	$\Delta$ & dead-band length & 2.5 & $^{\circ}$C			\\ 
	%\hline
	$P^i$ & Nominal power & 5.6 & kW\\ 
	%\hline
	$\eta$ & Coefficient of performance  & 0.3 & \\ \hline
	\end{tabular}
\end{center}
\end{table}
The TCLs are initialized at the steady state temperature condition $\theta^i[k]=\theta^i_{ref}$.
The reference signal is a normalized scaled down version of the frequency regulation signal from the PJM market~\cite{pjm_regulation_data} and is used to test the tracking performance of the stochastic battery model.
\subsection{Simulation Results}
\begin{figure} [htb]
\centering
% left, bottom, right and top 
\begin{overpic}[trim=0.2cm 0.5cm 0.4cm 0.5cm, width=0.48\textwidth]{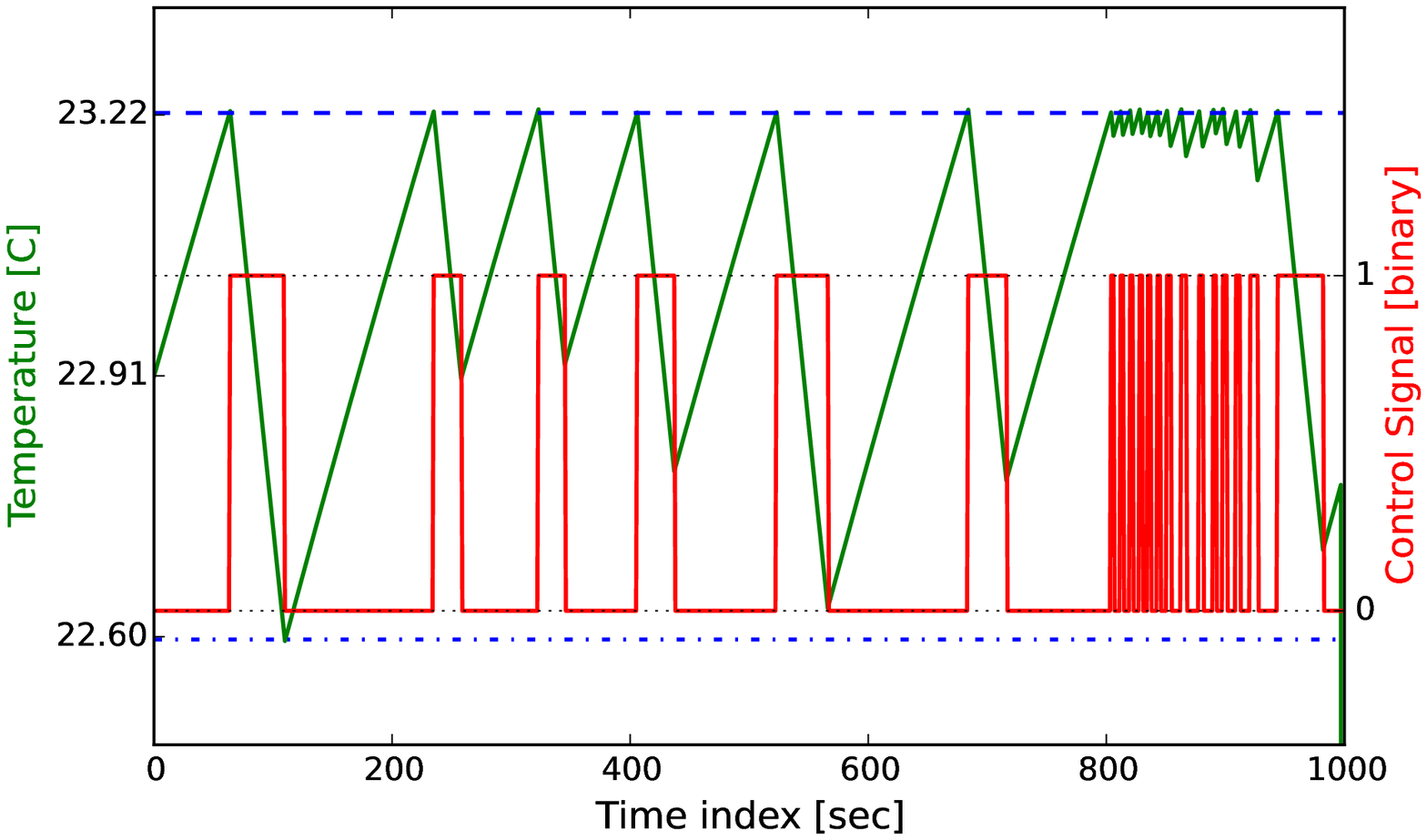}
\linethickness{1pt}
\put(76,48){\color{black}\line(1,0){12}}
\put (16,48)  {$\obar{\theta^i}$}
\put (16,6) {$\ubar{\theta^i}$}
\end{overpic}
\caption{TCL state transition dynamics (short-cycle duration: 2 sec)}
\label{figure:state transition of a tcl 2 sec}
\end{figure}

\begin{figure} [htb]
\centering
% left, bottom, right and top 
\begin{overpic}[trim=0.2cm 0.5cm 0.4cm 1cm, width=0.48\textwidth]{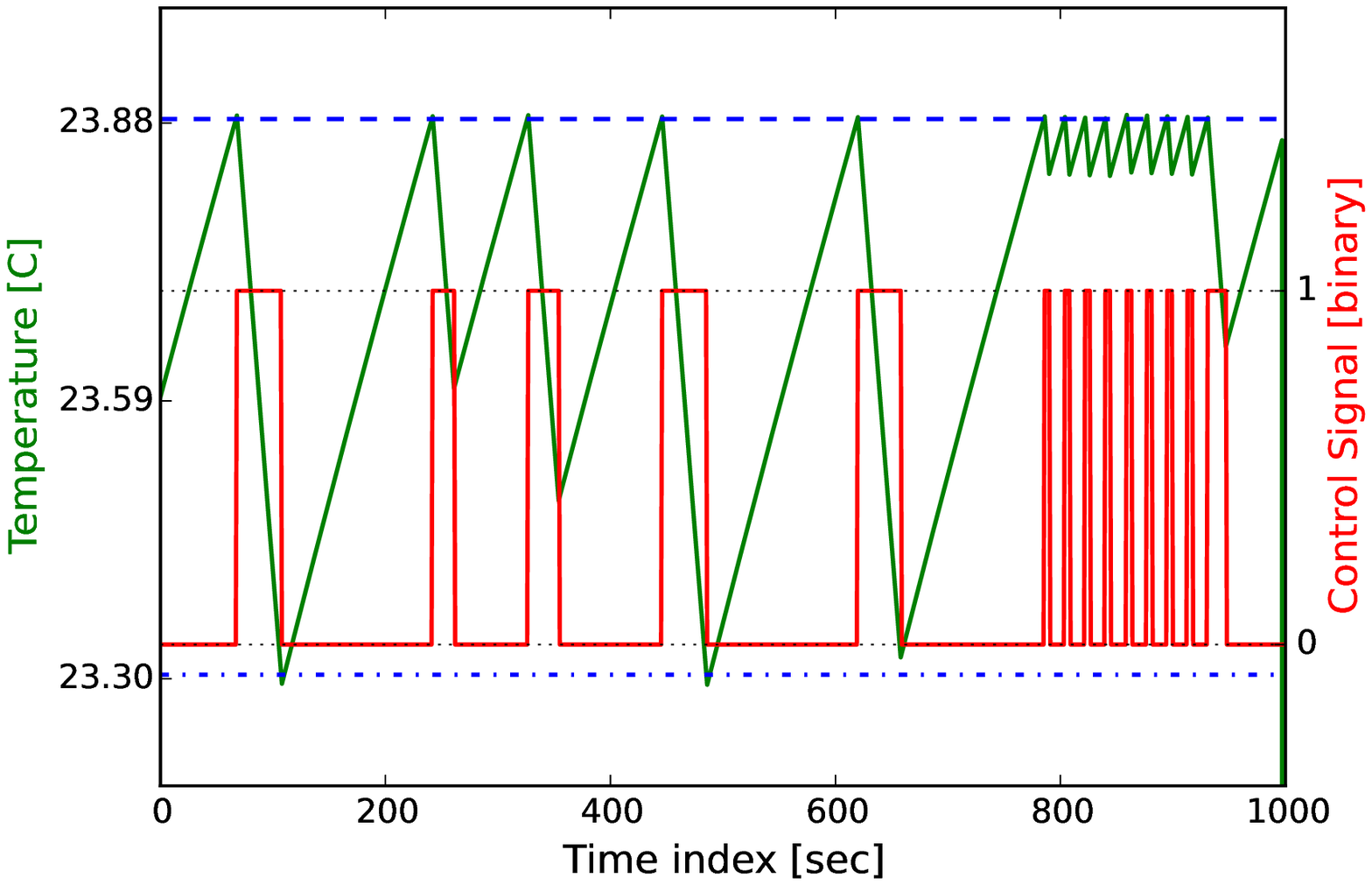}
\linethickness{1pt}
\put(74,53){\color{black}\line(1,0){13}}
\put (16,53)  {$\obar{\theta^i}$}
\put (16,7) {$\ubar{\theta^i}$}
\end{overpic}
\caption{TCL state transition dynamics (short-cycle duration: 6 sec)}
\label{figure:state transition of a tcl 6 sec}
\end{figure}

Fig.~\ref{figure:state transition of a tcl 2 sec} provides as insight into the temperature dynamics of the TCL when actively controlled externally.
It can be observed that the temperature evolves in dead-band and the state transition occurs at the boundaries.
While in between the temperature limits, the external signal can change the operational state provided the short cycling constraint is fulfilled.
Fig.~\ref{figure:state transition of a tcl 6 sec} shows the impact of the short cycling duration when increased from $2 sec.$ to $6 sec$. 
The short cycling duration values are selected for the proof of concept and can be conveniently changed to represent the actual requirements.
It is observed that when time is between $780$ to $950$ the TCL experiences repeated activation. 
This phenomenon is explained later on when the tracking of regulation signal is discussed.
%It is observed that when the short cycle duration is increased from $2$ to $6$ seconds the fluctuations decrease in the operation of the TCL as shown in Fig.~\ref{figure:state transition of a tcl 6 sec}.

\begin{figure} [htb]
\centering
% left, bottom, right and top 
\begin{overpic}[trim=0cm 0cm 0cm 1cm, width=0.48\textwidth]{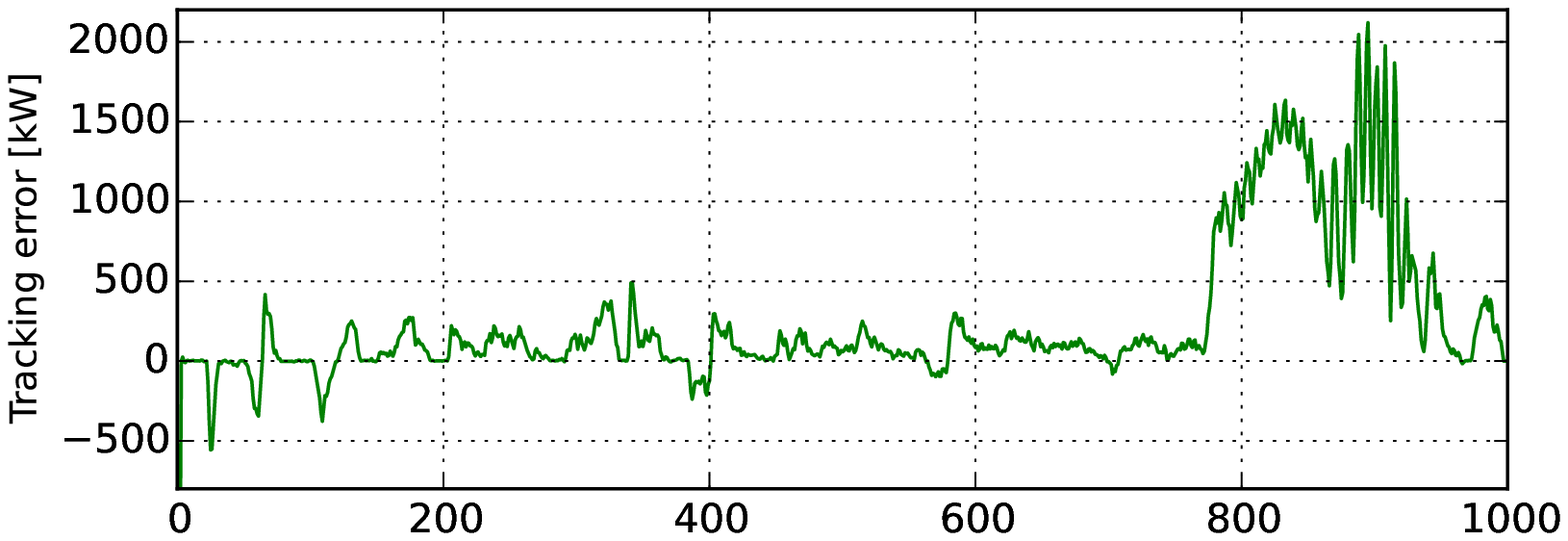}
\put (45,-3)  {\footnotesize$\textrm{Time [sec]}$}
\end{overpic}
\caption{Tracking error}
\label{figure:tracking error}
\end{figure}

\begin{figure} [htb]
\centering
% left, bottom, right and top 
\begin{overpic}[trim=0cm 0.5cm 0cm 0.5cm, width=0.5\textwidth]{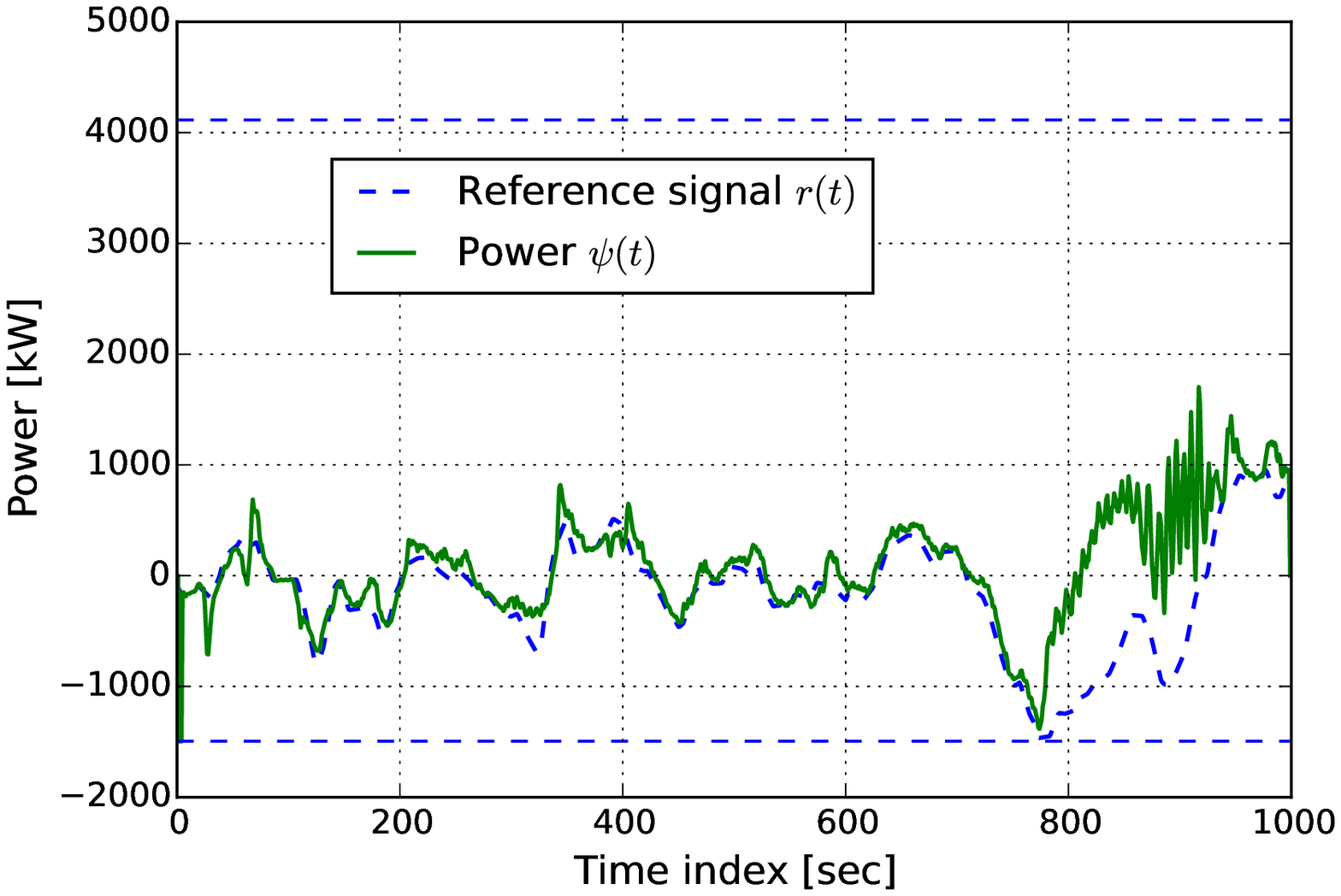}
\put (18,52)  {\footnotesize$R_+$}
\put (18,09) {\footnotesize$R_-$}
\end{overpic}
\caption{Tracking performance of the regulation signal}
\label{figure:tracking of regulation signal}
\end{figure}

\begin{figure} [htb]
\centering
% left, bottom, right and top 
\begin{overpic}[trim=0cm 0.5cm 0cm 0.5cm, width=0.5\textwidth]{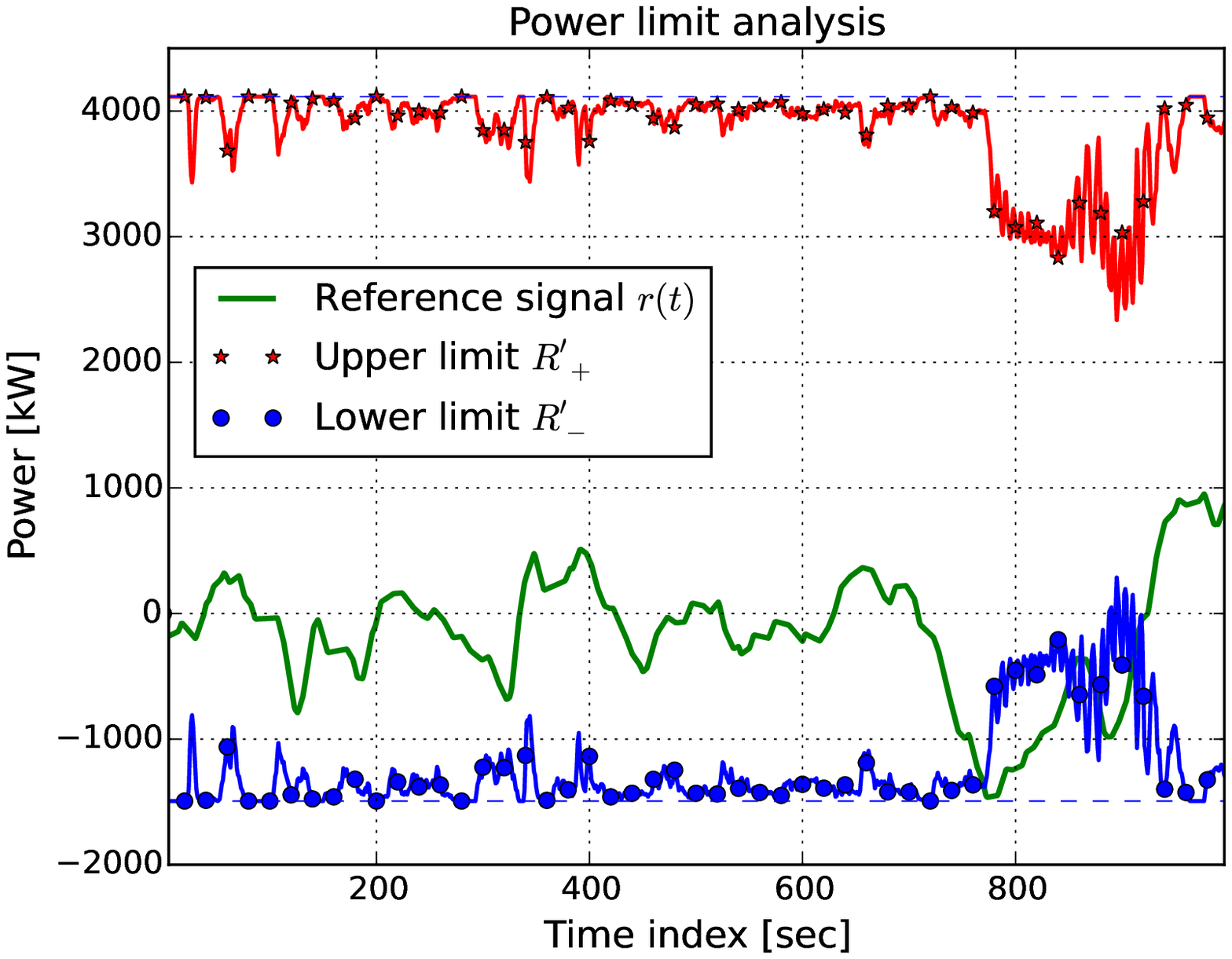}
\linethickness{1pt}
\put(17.1,42.1){\color{red}\line(1,0){2.5}}
\put(17.1,37.6){\color{blue}\line(1,0){2.5}}
%\put (18,62)  {\footnotesize$R_+$}
%\put (18,9) {\footnotesize$R_-$}
%\put (72,40)  {\footnotesize$R'_+$}
%\put (72,27) {\footnotesize$R'_-$}
\end{overpic}
\caption{Tracking performance comparison with stochastic regulation limits}
\label{figure:stochastic regulation limits}
\end{figure}

\begin{figure} [htb]
\centering
% left, bottom, right and top 
\begin{overpic}[trim=0cm 0.5cm 0cm 0.5cm, width=0.5\textwidth]{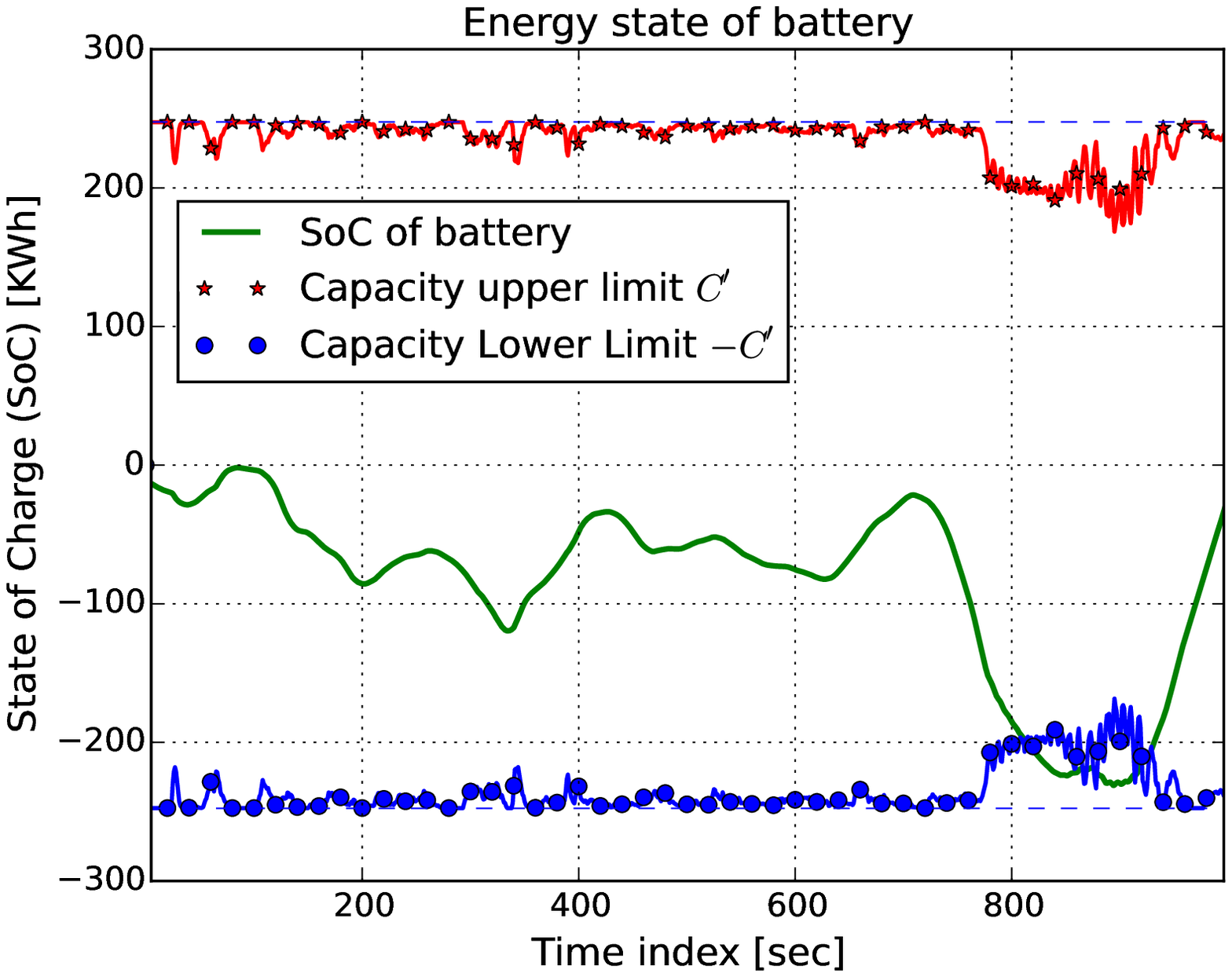}
\linethickness{1pt}
\put(17.1,47.7){\color{red}\line(1,0){2.5}}
\put(17.1,43.5){\color{blue}\line(1,0){2.5}}
%\put (76,36)  {\footnotesize$C$}
%\put (76,16) {\footnotesize$-C$}
%\put (18,64)  {\footnotesize$C$}
%\put (18,9) {\footnotesize$-C$}
\end{overpic}
\caption{State of charge dynamics of the stochastic battery model}
\label{figure:state of charge dynamics}
\end{figure}

\begin{figure} [htb]
\centering
% left, bottom, right and top 
\begin{overpic}[trim=0cm -0.2cm 0cm 0.5cm, width=0.5\textwidth]{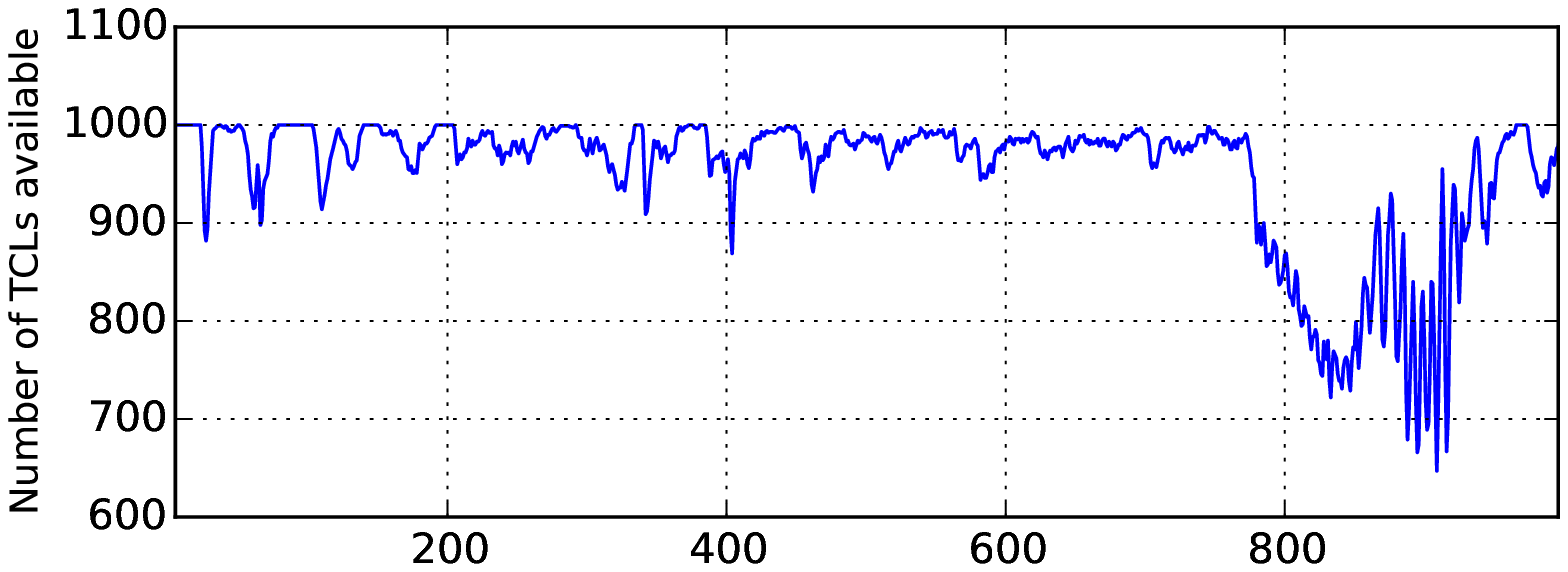}
\put (45,-2)  {\footnotesize$\textrm{Time [sec]}$}
%\Huge
%\huge
%\LARGE
%\large
%\normalsize (default)
%\small
%\footnotesize
%\scriptsize
%\tiny
\end{overpic}
\caption{TCL availability dynamics}
\label{figure:tcl availability}
\end{figure}

The repeated activation phenomenon can be observed by analysis of the regulation signal dynamics.
The tracking performance for the test case is shown in Fig.~\ref{figure:tracking of regulation signal}.
The TCLs aggregation tracks the regulation signal till $k=780\;\textrm{sec}$.
The tracking is lost onward despite the reference signal occurring within the regulation bounds.
The reason behind is that the reference signal violates the stochastic limit constraints.
This aspect is captured by dynamic limits in Eq.~\ref{eq:stochastic battery model}.
The tracking error as the result of violation of this constraint can be seen in \fig~\ref{figure:tracking error}.
The stochastic regulation limits for this case are shown in Fig.\ref{figure:stochastic regulation limits}.
It can be observed that the reference signal violates the stochastic ramp-rate limits.
The reasons behind the decrease in the permissible ramp-rate limits is the the unavailability of the TCLs shown in Fig.~\ref{figure:tcl availability}.
Furthermore, the stochastic capacity limit of the battery given as $C'$ also changes with time. The dynamics are shown in Fig.~\ref{figure:state of charge dynamics}. 
The formulation provides dynamic bounds on these variables.
The tracking of the regulation signal is ensured if the stochastic SoC and ramp-rate limits are fulfilled.

\subsection{Outlook}
The results shows that stochastic modeling of resources during operation shall be an important aspect of demand side management.
In order to ensure the regulatory requirements, the need to actively control the resources is inevitable~\cite{6832599}.
The proposed formulation can be used to actively plan alternate resources based on the state of the stochastic battery model.
The dynamics of the regulation signal can be filtered based on the proposed battery model and the required flexibility can be allocated to other resources. 
A mechanism proposed in~\cite{7346488} can be used for this purpose.
Apart from the direct control of the loads, the price based and incentive based approach can be integrated in this framework as well.
In this case the price or incentives can alter the behavior of the TCL loads.% and hence adjustments can be made during operation.
The information about the stochastic battery model parameters thus provide a reference to facilitate operation under uncertainty.
In case of the uncertain short cycling duration and communication delay, a joint probability distribution can be used to predict the tracking accuracy while using this model.

\section{Conclusion}\label{sec:conclusion} 
This paper presents an novel approach that builds upon the existing literature for calculating the stochastic parameters of battery representing the TCL aggregation.
The parameters of stochastic state of charge and capacity limits are computed while considering resources availability directly.
The result are the probabilistic bounds on the regulation signal dynamics that if observed guarantee the tracking.
The presented approach prevents the calculation of the power at each TCL thus decreasing the operational cost. 
Alternatively, it is proposed here that the TCL ratings is provided in form of the contract with the central control. 
The availability signal is communicated by the TCL to represent its current state.
A TCL is required to communicate the availability, status and relative temperature distance to switching boundary which can be achieved with a nominal bandwidth.


%\section*{Acknowledgment}



%\bibliographystyle{IEEEtran}
%\begin{thebibliography}{1}
%\end{thebibliography}

\end{document}